# Integrated Solution Modeling Software: A New Paradigm on Information Security Review and Assessment

Heru Susanto[123], Mohammad Nabil Almunawar[1], Yong Chee Tuan[1],
Mehmet Sabih Aksoy[3] and Wahyudin P Syam[4]

[1]**FBEPS University of Brunei**
Information System Group
susanto.net@gmail.com

[2]**The Indonesian Institute of Sciences**
Information Security &
IT Governance Research Group
heru.susanto@lipi.go.id

[3]**King Saud University**
Information System Department
hsusanto@ksu.edu.sa

[4]**Politecnico**
di' Milano

*Abstract*— Actually Information security becomes a very important part for the organization's intangible assets, so level of confidence and stakeholder trusted are performance indicator as successes organization. Since information security has a very important role in supporting the activities of the organization, we need a standard or benchmark which regulates governance over information security. The main objective of this paper is to implement a novel practical approach framework to the development of information security management system (ISMS) assessment and monitoring software, called by I-SolFramework. System / software is expected to assist stakeholders in assessing the level of their ISO27001 compliance readiness, the software could help stakeholders understood security control or called by compliance parameters, being shorter and more structured. The case study illustrated provided to the reader with a set of guidelines, that aims easy understood and applicable as measuring tools for ISMS standards (ISO27001) compliance.

|  | ISBS 2010 small organisations | ISBS 2010 large organisations |
|---|---|---|
| Business disruption | £15,000 - £30,000 over 2-4 days | £200,000 - £380,000 over 2-5 days |
| Time spent responding to incident | £600 - £1,500 2-5 man-days | £6,000 - £12,000 15-30 man-days |
| Direct cash spent responding to incident | £4,000 - £7,000 | £25,000 - £40,000 |
| Direct financial loss (e.g. loss of assets, fines etc.) | £3,000 - £5,000 | £25,000 - £40,000 |
| Indirect financial loss (e.g. theft of intellectual property) | £5,000 - £10,000 | £15,000 - £20,000 |
| Damage to reputation | £100 - £1,000 | £15,000 - £200,000 |
| **Total cost of worst incident on average** | **£27,500 - £55,000** | **£280,000 - £690,000** |
| 2008 comparative | £10,000 - £20,000 | £90,000 - £170,000 |

*Table 1*. The overall cost of an organization's worst incident

## I. INTRODUCTION

Today, security is very hot issue and topic to be discussed, ranging from business activities, correspondence, banking and financial activities are crucial, it requires prudence and high precision. Recent news indicates how many cases of data theft and credit card pishing in the information that led to enormous losses, so that someone will get a large bill while in question was not using the credit card. Information is the lifeblood of organizations, a vital business asset in today's IT-enabled world. Access to high-quality, complete, accurate and up-to-date information is vital in supporting managerial decision-making process that leads to sound decisions. Thus, securing information system resources is extremely important to ensure that the resources are well protected.

*Table 1* showed us the total cost of incident increases every year, as indicated by information security breaches survey (*Chris Potter & Andrew Beard, 2010*), this phenomenon is the negative effect for the operation of an organization. Research on information security area is extremely needed, especially in order to deal with the connectivity and cloud computing era. Information security is not just a simple matter of having usernames and passwords (*Alan Calder and Setve Watkins, 2008*). Regulations and various privacy/data protection policy impose a raft of obligations to organization. Meanwhile, viruses, worms, hackers, phishers and social engineers threaten an organization on all sides

Although the development of IT security framework has gained much needed momentum in recent years, there continues to be a need for more writings on best theoretical and practical approaches to security framework development. Thus, securing information system resources is extremely important to ensure that the resources are well protected (*Chris Potter & Andrew Beard, 2008*). Regulations and various privacy/data protection policy impose a raft of obligations to organization. Meanwhile, viruses, worms, hackers, phishers and social engineers threaten an organization on all sides. Organizational communication channels, which use network technology, intranet, extranet, internet, are a target for hackers in filtrated by [*figure 1*].

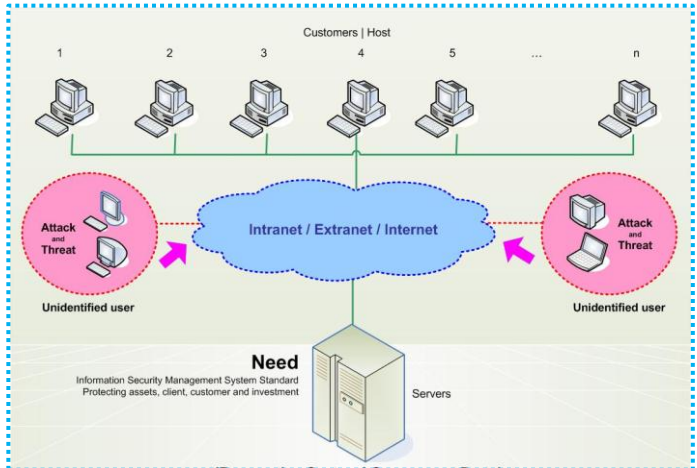

*Figure 1*. Activities of unidentified user as potential attack and threat to organization

Furthermore, comprehensive and reliable information security controls reduce the organization's overall risk profile. ISO 27001 is the standard relating to Information Security Management System (ISMS). Companies or organizations obtained of ISO 27001 Certificate meaning a well recognized for the security of information systems. Meaning, with this certification, the company's credibility is expected to get good recognition from customers and client, thereby increasing the level of trust and the number of customers, which in turn will increase the profit of the company.

Since information security has a very important role in supporting the activities of the organization, we need a standard as benchmark which regulates governance over information security. Several private and government organizations developed standards bodies whose function is to setup benchmarks, standards and in some cases, legal regulations on information security to ensure that an adequate level of security is preserved, to ensure resources used in the right way, and to ensure the best security practices adopted in an organization. There are several standards for IT Governance which leads to information security such as PRINCE2, OPM3, CMMI, P-CMM, PMMM, ISO27001, BS7799, PCIDSS, COSO, SOA, ITIL and COBIT. However, some of these standards are not well adopted by the organizations, with a variety of reasons. In this paper we will discuss the big five of ISMS standards, widely used standards for information security. The big five are ISO27001, BS 7799, PCIDSS, ITIL and COBIT. The comparative study conducted to determine their respective strengths, focus, main components and their adoption based on ISMS (*susanto, almunawar & yong, 2011b*) concluded that ISO 27011 is most widely used standard in the world in information security area.

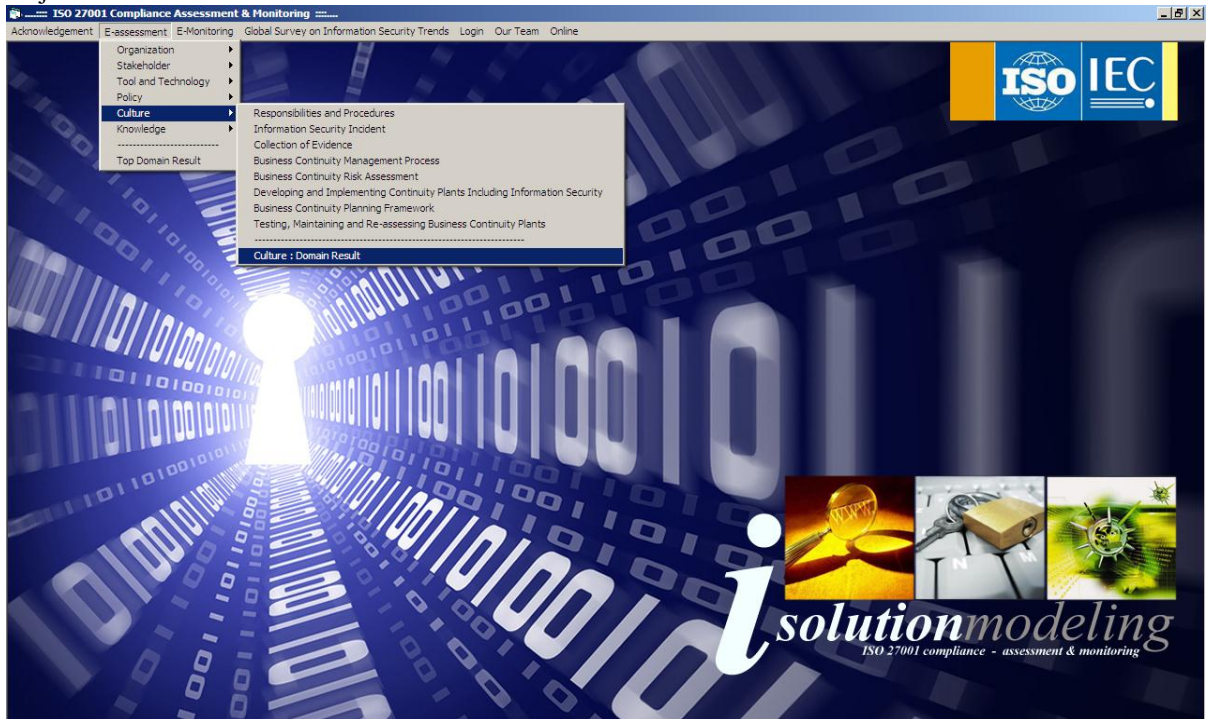

*Figure 2*. Integrated solution modelling software view

Unfortunately, many organizations find it difficult to implement ISO 27001, including the obstacle when measuring the level of organizational readiness, which includes preparation of documents as well as various scenarios relating to information security (*susanto, almunawar & yong, 2011a*) and (*siponen & willison. 2009*). In addressing issue mentioned this research will provide creative solutions and a new paradigm in measuring readiness level of ISO27001 compliance, by developing an application system / software called by integrated modeling solution for ISO 27001 assessment and monitoring compliance (i-Solution Modeling sofwtare) [*figure 2*].

## II.    I-SOLUTION FRAMEWORK

This section we introduced new framework for approaching object and organization analyst, called by I-SolFramework, abbreviation from *I*ntegrated *Sol*ution for Information Security *Framework*. The framework consists of six layers component *[figure 3]*: organization, stakeholder, tools & technology, policy, culture, knowledge. We introduced the basic elements of development in I-SolFramework profile as illustrated (*susanto, almunawar & yong, 2011c*) & (*alfantookh, 2009*).

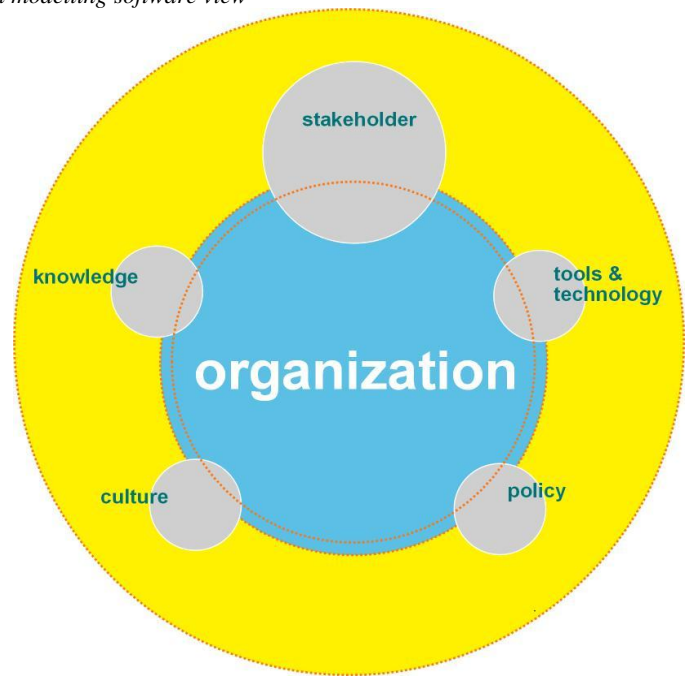

*Figure 3*. Integrated solution six domain framework

## III. ELUCIDATION OF TERM AND CONCEPT

|  | ISO 27001 Issues | Controls | |
|---|---|---|---|
|  |  | Essential | |
|  |  | Section | No. |
| **Policy** | Information Security Policy | 5.1.1 | 1 |
| **Tool &Technology** | Information Systems Acquisition, Development and Maintenance | 12.2.1<br>12.2.2<br>12.2.3<br>12.2.4<br>12.6.1 | 5 |
| **Organization** | Organization of Information Security | 6.1.3 | 1 |
| **Culture** | Information Security Incident Management | 13.2.1<br>13.2.2<br>13.2.3 | 3 |
| | Business Continuity Management | 14.1.1<br>14.1.2<br>14.1.3<br>14.1.4<br>14.1.5 | 5 |
| **Stakeholder** | Human Resources Security | 8.2.1<br>8.2.2<br>8.2.3 | 3 |
| **Knowledge** | Compliance | 15.1.2<br>15.1.3<br>15.1.4 | 3 |
| **Total objectives and controls** | | | 21 |

*Table 2. A view of ISO 27001 clauses, objectives, controls and essential controls*

1. **Organization:** A social unit of people, systematically structured and managed to meet a need or to pursue collective goals on a continuing basis, the organizations associated with or related to, the industry or the service concerned (*Business Dictionary*).

2. **Stakeholder:** A person, group, or organization that has direct or indirect stake in an organization because it can affect or be affected by the organization's actions, objectives, and policies (*Business Dictionary*).

3. **Tools & Technology:** the technology upon which the industry or the service concerned is based. The purposeful application of information in the design, production, and utilization of goods and services, and in the organization of human activities, divided into two categories (1) Tangible: blueprints, models, operating manuals, prototypes. (2) Intangible: consultancy, problem-solving, and training methods (*Business Dictionary*).

4. **Policy:** typically described as a principle or rule to guide decisions and achieve rational outcome(s), the policy of the country with regards to the future development of the industry or the service concerned (*Business Dictionary*).

5. **Culture:** determines what is acceptable or unacceptable, important or unimportant, right or wrong, workable or unworkable. ***Organization Culture:*** The values and behaviors that contribute to the unique social and psychological environment of an organization, its culture is the sum total of an organization's past and current assumptions (*Business Dictionary*).

6. **Knowledge:** in an organizational context, knowledge is the sum of what is known and resides in the intelligence and the competence of people. In recent years, knowledge has come to be recognized as a factor of production (*Business Dictionary*).

## IV. MATHEMATICAL MODELS

The mathematical model is explaining us on mentioned algorithms as well as facilitating readers gain a comprehensive and systematic overview of the mathematical point of view. Modeling begins by calculating the lowest level components of the framework, namely *section*. Determining the lowest level could be flexible, depending on the problems facing the object, might be up to $3^{rd}$, $4^{th}$, $5^{th}$ ... $N^{th}$ level.

Formula works recursively; enumerate value from the lowest level, until the highest level of framework. several variables used as position and contents of framework indicator, where $k$ as *control*, $j$ as *section*, and $h$ as a *top level*, details of these models are mention as follows.

$$(a) \to x_j = \sum_{k=1}^{n} \frac{[section]_k}{n}$$

$$x_j: control$$

$(a) \to x_j$ Indicate value of control of ISO which is resulting from *sigma* of section(s) assessment, divided by number of section(s) contained on the lowest level.

$$(b) \to x_i = \sum_{j=1}^{n} \frac{[control]_j}{n}$$

$$x_i: domain$$

$(b) \to x_i$ Stated value of domain of ISO which is resulting from *sigma* of control(s) assessment, divided by number of control(s) contained at concerned level.

After *(a), (b),* well defined, then the next step is substituted of mathematical equations mentioned above, into new comprehensive modeling notation in a single mathematical equation, as follows.

$$x_h = \sum_{i=1}^{n} \frac{[control]_i}{n}$$

$$x_h = \sum_{i=1}^{n} \frac{[b]_i}{n}$$

$$x_h = \sum_{i=1}^{n} \frac{\left[\sum_{j=1}^{n} \frac{[section]_j}{n}\right]_i}{n}$$

*So that for six layer, or we called it by top level, equation will be:*

$$x_h = \sum_{i=1}^{6} \frac{\left[\sum_{j=1}^{n} \frac{\left[\sum_{k=1}^{n} \frac{[section]_k}{n}\right]_j}{n}\right]_i}{6}$$

*Where; k=section; J=control; I=domain (organization, stakeholder, tools & technology, policy, knowledge, and culture).*

The algorithm is considered to be reliable and easy implementing in analyzing such problem (*susanto, almunawar & tuan, 2011d*), emphasized on divided problems into six layers as the initial reference in measuring and analyzing the object. The results obtained as full figure indicators of an organization's readiness in information security compliance, it showed us strong and weak point on layer of the object. Indicated that layer with a weak indicator has a high priority for improvement and refinement within organization as a whole. In the manner of the six layer framework, I-SolFramework, analysis could be done easily and simply observe.

## V. SOFTWARE DESIGN

The ISO recommendations are associated with two levels of security protection: a basic level that considers essential security controls; and an extended level that extends the essential controls in order to provide additional security protection (*Kosutic, 2010*). It should be noted here that some organizations may not only consider what ISO/IEC recommend, but they may also add to them special controls needed for the protection of their work, in order to achieve their business objectives (*Alfantookh, 2009*). It starts with the "21 essential security controls" of ISO 27001, which give the basic standard requirements of information security management. The controls are mapped on these domains and subsequently refined into "246 simple and easily comprehended elements". These elements are subject to be reviewed and validated by specialized persons working on the field.

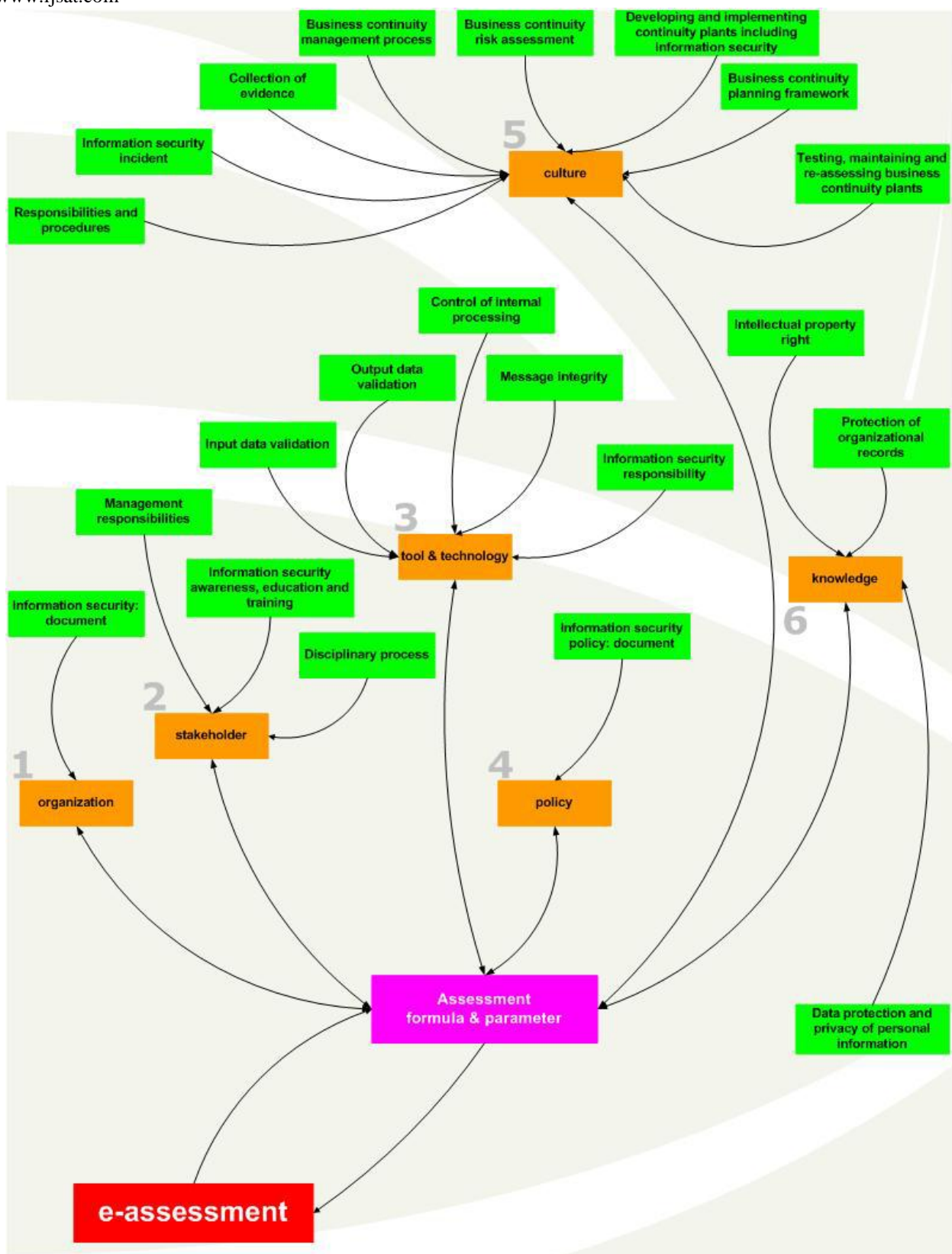

*Figure 4*. I-Solution Modeling Data flow diagram level 1

In general i-Solution Modeling consists of two major subsystems of e-assessment and e-monitoring [*figure 5*]. E-assessment to measure ISO 27001 parameters [*Table 2*] based on the proposed framework [*figure 3*] with 21 controls [*figure 4*]. Software is equipped with a login system, as the track record of the user, how many times try the experiment in assessing the organization so it can be determined patterns of assessment. Database will be updated automatically, so query and retrieval of newest records will be neater structured and well organized.

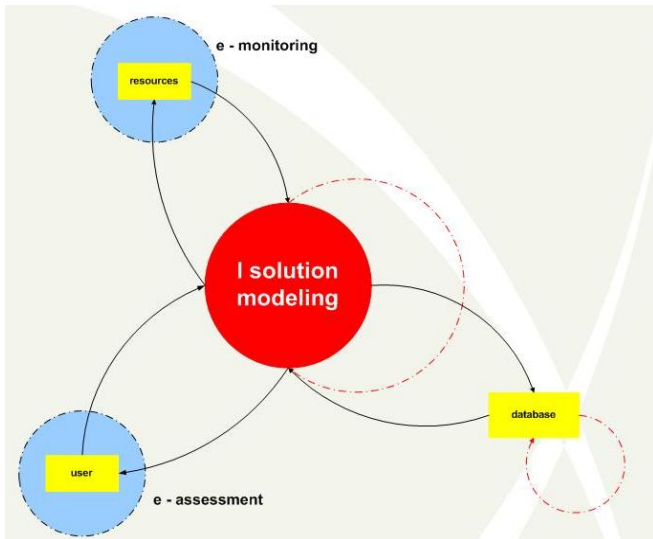

*Figure 5*. I-Solution Modeling Data flow diagram level 0

E-assessment part is validating of the ISO 27001 parameters, by the user interface provided by the system, follows rules of i-solution framework rule that divided and segmented ISO 27001 essential controls into six main domains [*figure 4*].

Stakeholders have to inserting grade level as implementation measurement value of the implemented security parameters. Level of assessment set out in range of 5 scales;

- ❖ *0 = not implementing*
- ❖ *1= below average*
- ❖ *2=average*
- ❖ *3=above average*
- ❖ *4=excellent*

## VI. SOFTWARE FEATURES

In the main form, software is featuring by three main tabs as function as:

- *Tab Assessment*
  On this sub form, user is prompted to entering an achievements value they done based on ISO 27001 parameters, named as Assessment issues. Level of assessment set out in range of 5 scales;
  - ❖ *0 = not implementing*
  - ❖ *1= below average*
  - ❖ *2=average*
  - ❖ *3=above average*
  - ❖ *4=excellent*

As a measurement example for the parameters bellows [*figure 6*]:
  1. *Domain*: "**Organization**"
  2. *Controls*: "**Organization of information security: Allocation of Information Security Responsibilities**"
  3. *Assessment Issue*: "**Are assets and security process Cleary Identified?**"
  4. Then stakeholders should analogize ongoing situation, implementation and scenario in organization, and benchmark it to the security standard level of assessment as reference standard.

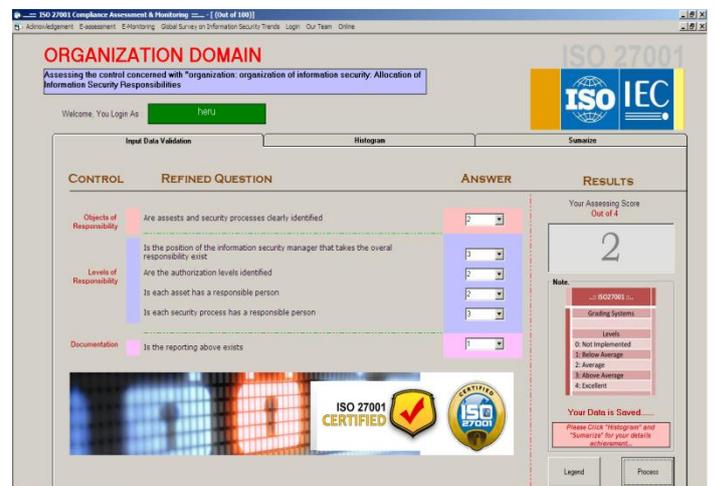

*Figure 6*. Assessment form

- *Tab Histogram*
  Histogram showed us details of the organization's achievement and priority. Both statuses are important in order to review of strongest points and weakness point.

  *"Achievement"* declared the performance of an organization which is the result of the measurement by the framework.

  *"Priority"* indicated the gap between ideal values with achievement value. "Priority" and "achievement" it showed inverse relationship. If achievement is high, then domain has a low priority for further work, and conversely, if achievement is low, then the priority will be high [*figure 7*].

Priority status or it could be referred as scale of priorities, is the best reference for evaluating of the organization based on six main domains. Scale of priority value helping us on evaluating, auditing and maintaining a problem, its becomes easier and highly precise, since the stakeholders and users do not need to evaluate all components of the organization, but could focused on repairs and improvements in the domain with low achievement grade.

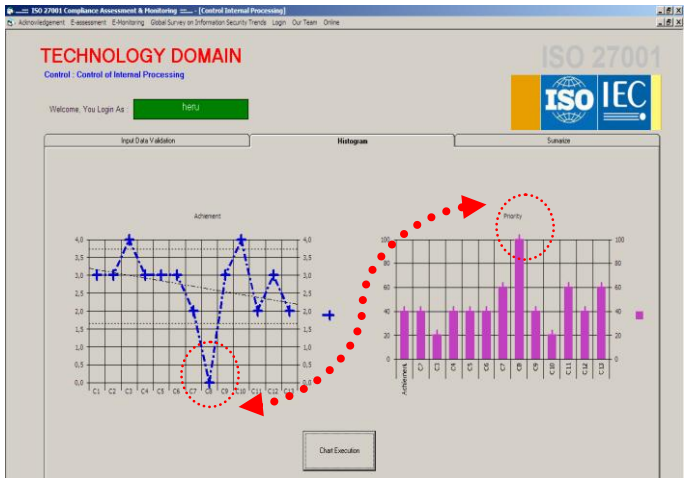

*Figure 7. Final result view on histogram style*

- ***Tab Summarize***

    Sub menu summarize has a feature that provided a user to analyze the results of their estimates. Some of the assessment criteria are displayed, in addition system advice is prepared, and it gives advice based on the previous assessment. Four main features are:

    ❖ Final result out of 4 scale
    ❖ Final result out of 100 %
    ❖ Final predicate of assessment result (not implementing, below average, average, above average, excellent)
    ❖ Advice from the software regarding their final achievement, in which point their strongest area and also their weakness area.

    By marking estimated performance values for each parameter, as assessment and forecasting approach as well, stakeholders have a comprehensive overview achievement on their readiness level. In some cases, to assess their organization's readiness level, stakeholders needed more than five times of experiment, marked by a significant increase in the final result grade of each experiment. Two first experiments marking on stakeholder assessment called by first trying and training course, time required for each experiment was 30 minutes to 60 minutes, this achievement represents a significant contribution to an organization in understanding the ISO 27001 controls, clause and assessment issues as well, than normally step which is need. 12 months - 24 months (*iso27001security.com*).

## VII. AN ILLUSTRATIVE MEASUREMENT

An illustrative example is presented to delineate usability level of its approach. Each question of the refined simple elements, a value associated with the example is given. *Table 2* summarizes the results of all domains together with their associated controls (*susanto, almunawar & yong, 2011c*) & (*alfantookh, 2009*), based on I-SolFramework measurements approaches.

*Figure 8. Final result view on summarize style*

The results given are illustrated in the following Figures. *Figure 9* illustrates the state of five essential controls of the "tools & technology" domain. *Figure 8* represents the condition of 21 essential controls of standards by table and also *Figure 10* stated overall condition of 21 essential controls in histogram style.

The overall score of all domains is shown in the Table to be "2.66 points". The domain of the "policy" scored highest at "4", and the domain of the "knowledge" scored lowest at "2". Ideal and priority figures are given to illustrate the strongest and weaknesses in the application of each control (*figure 9 &10*).

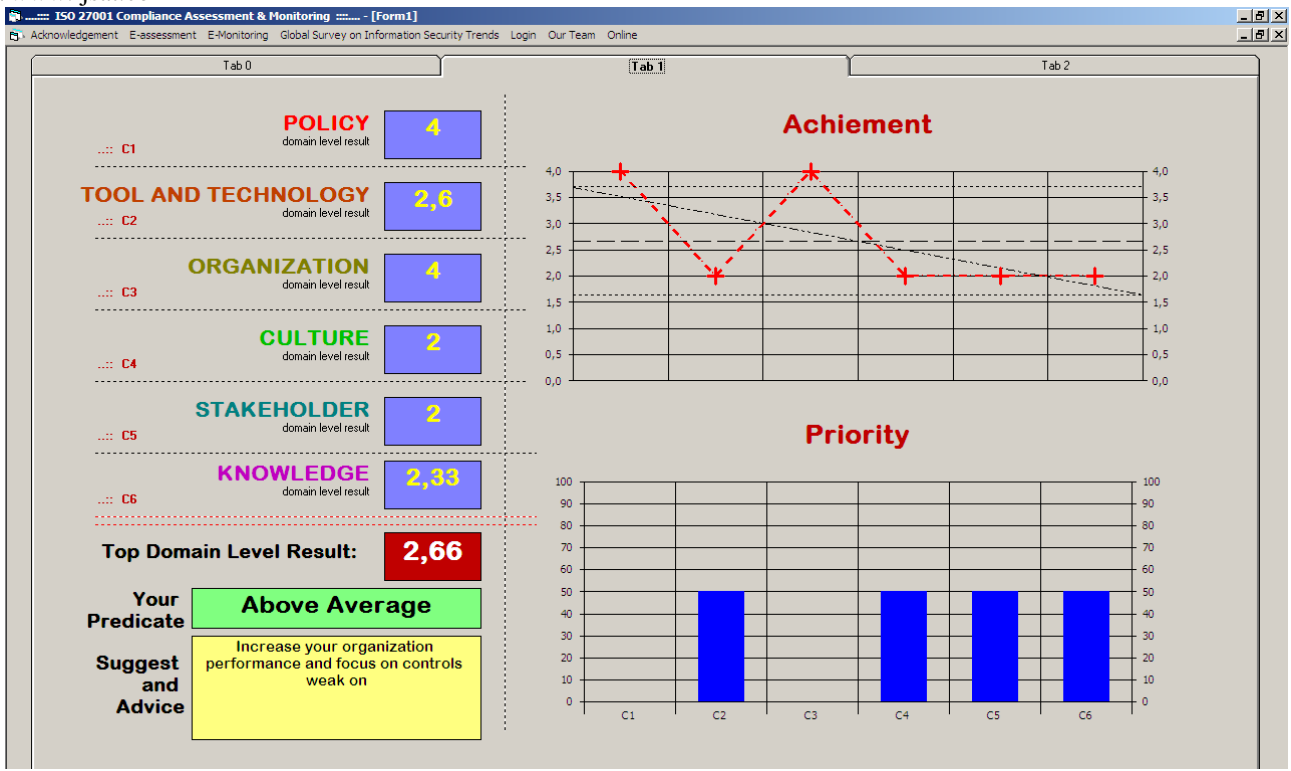

*Figure 9*. Six domain final result view on histogram style

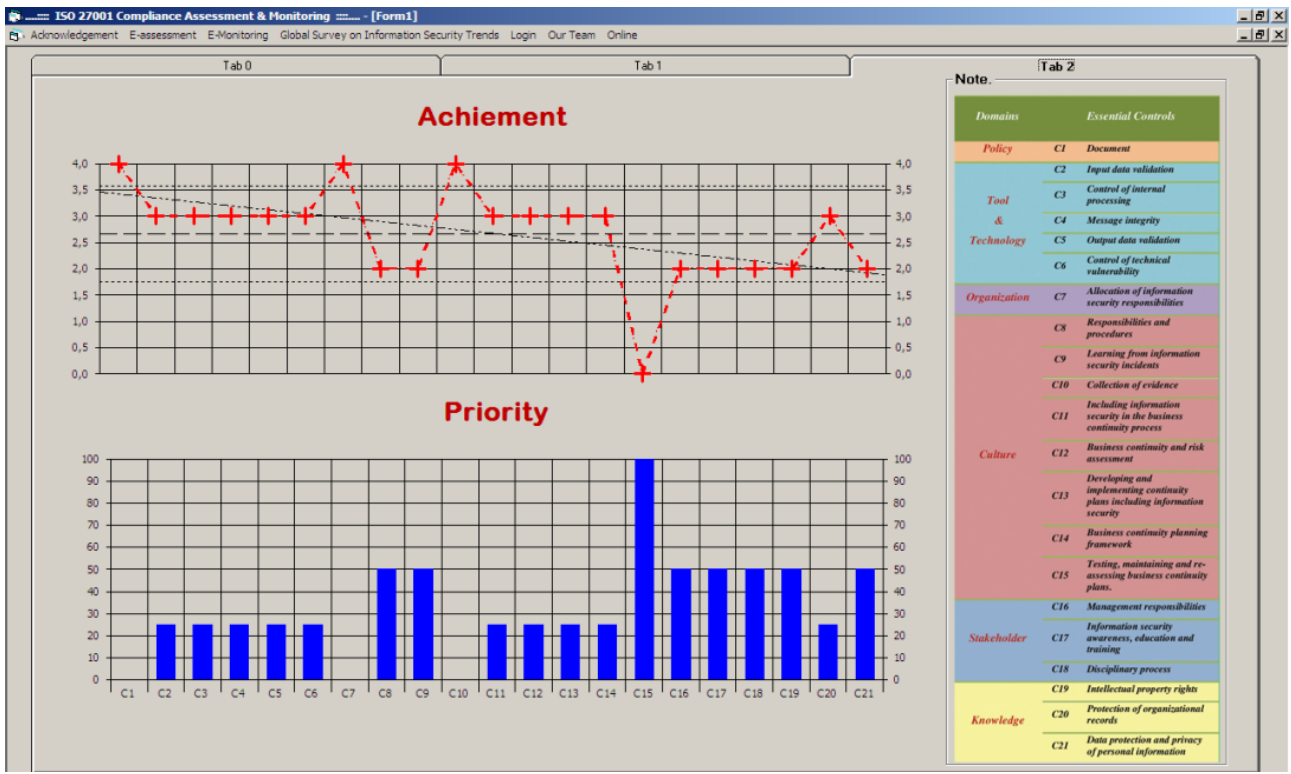

*Figure 10*. 21-essential controls final result view on histogram style

## VIII. Conclusion Remarks

Securing and maintaining information from parties who do not have authorization to access such information is crucial priority. It is important for an organization to implement an information security standard, ISO 27001 as a reference. I-Solution modeling is software which has new paradigm framework, to make assessments and monitoring. It is expected to provide solutions to solved obstacles, challenges and difficulties in understanding standard term and concept, as well as assessing readiness level of an organization towards implementation of ISO 27001 for information security. On trials conducted, user can perform a test of his organization assessment within 30-60 minutes; in expected perhaps it could reduce time in understanding and assessing set of standard parameters leads obtain certification of information security.

## IX. References

Abdulkader Alfantookh. An Approach for the Assessment of The Application of ISO 27001 Essential Information Security Controls. Computer Sciences, King Saud University. 2009.

Alan Calder and Setve Watkins. *IT Governance – A Manager's Guide to Data Security and ISO 27001 and ISO 27002*. 2008.

Chris Potter & Andrew Beard. *Information Security Breaches Survey 2010*. Price Water House Coopers.Earl's Court,London.2010.

Dejan Kosutic. 2010. ISO 27001 and BS 25999. Obtained from http://blog.iso27001standard.com

Heru Susanto & Fahad bin Muhaya. *Multimedia Information Security Architecture*. @IEEE. 2010.

Heru Susanto, Mohammad Nabil Almunawar & Yong Chee Tuan. *I-SolFramework View on ISO 27001. Information Security Management System: Refinement Integrated Solution's Six Domains*. Asian Transaction on Computer Journal. 2011a.

Heru Susanto, Mohammad Nabil Almunawar & Yong Chee Tuan. *Information Security Management System Standards: A Comparative Study of the Big Five*. International Journal of Engineering and Computer Science. 2011b.

Heru Susanto, Mohammad Nabil Almunawar & Yong Chee Tuan. *An Integrated Solution Framework a Tool forMeasurement and Refinement of Information Security Management Standard*. On reviewed paper. 2011c.

Heru Susanto, Mohammad Nabil Almunawar & Yong Chee Tuan. *I-SolFramework: An Integrated Solution Framework Six Layers Assessment on Multimedia Information Security Architecture Policy Compliance*. Advanced in Multimedia Journal. Hindawi Publisher. On reviewed. 2011d.

Mikko Siponen & Robert Willison. 2009. *Information securitystandards: Problems and Solution*. Information & Management 46(2009) 267-270. Elsevier Science Ltd.

Web Site Business Dictionary online, over 20.000 terms. *www.businessdictionary.com*

www.iso27001security.com

## Authors

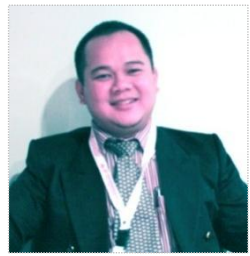

**Heru Susanto** is a researcher at The Indonesian Institute of Sciences, Information Security & IT Governance Research Group, also was working at Prince Muqrin Chair for Information Security Technologies, Information Security Research Group, King Saud University. He received BSc in Computer Science from Bogor Agriculture University, in 1999 and MSc in Computer Science from King Saud University, and nowadays as a PhD Candidate in Information System from the University of Brunei.

**Mohammad Nabil Almunawar** is a senior lecturer at Faculty of Business, Economics and Policy Studies, University of Brunei Darussalam. He received master Degree (MSc Computer Science) from the Department of Computer Science, University of Western Ontario, Canada in 1991 and PhD from the University of New South Wales (School of Computer Science and Engineering, UNSW) in 1997. Dr Nabil has published many papers in refereed journals as well as international conferences. He has many years teaching experiences in the area computer and information systems.

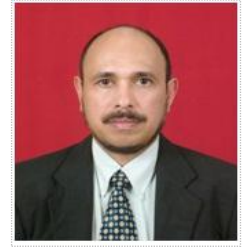

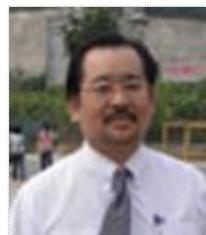

**Yong Chee Tuan** is a senior lecturer at Faculty of Business, Economics and Policy Studies, University of Brunei Darussalam, has more than 20 years of experience in IT, HRD, e-gov, environmental management and project management. He received PhD in Computer Science from University of Leeds, UK, in 1994. He was involved in the drafting of the two APEC SME Business Forums Recommendations held in Brunei and Shanghai. He sat in the E-gov Strategic, Policy and Coordinating Group from 2003-2007. He is the vice-chair of the Asia Oceanic Software Park Alliance.

**Mehmet Sabih Aksoy** is a Professor in Information System, King Saud Univrsity. He received BSc from Istanbul Technical University Engineering Faculty Dept. of Mechanical Engineering Istanbul 1982. MSc from Yildiz University Institute of Science Dept. of Industrial Engineering Istanbul 1985 and PhD from University of Wales College of Cardiff Electrical, Electronic and Systems Engineering South Wales UK 1994. Prof Aksoy interest on several area such as Machine learning, Expert system, Knowledge Acquisition, Computer Programming, Data structures and algorithms, Data Mining, Artificial Neural Networks, Computer Vision, Robotics, Automated Visual inspection, Project Management.

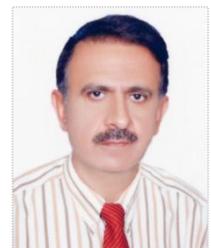

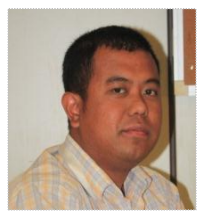

**Wahyudin P Syams** was receiving BSc from University of Indonesia and MSc from King Saud University, in the area of industrial engineering, actually now he was a PhD candidate from Politecnico di' Milano. Wahyudin interests on modeling and prototyping especially in manufacturing area.